# Academic Collaboration Rates and Citation Associations Vary Substantially Between Countries and Fields[1]

Mike Thelwall, Nabeil Maflahi
Statistical Cybermetrics Research Group, University of Wolverhampton, UK.

Research collaboration is promoted by governments and research funders but if the relative prevalence and merits of collaboration vary internationally different national and disciplinary strategies may be needed to promote it. This study compares the team size and field normalised citation impact of research across all 27 Scopus broad fields in the ten countries with the most journal articles indexed in Scopus 2008-2012. The results show that team size varies substantially by discipline and country, with Japan (4.2) having two thirds more authors per article than the UK (2.5). Solo authorship is rare in China (4%) but common in the UK (27%). Whilst increasing team size associates with higher citation impact in almost all countries and fields, this association is much weaker in China than elsewhere. There are also field differences in the association between citation impact and collaboration. For example, larger team sizes in the Business, Management & Accounting category do not seem to associate with greater research impact, and for China and India, solo authorship associates with higher citation impact. Overall, there are substantial international and field differences in the extent to which researchers collaborate and the extent to which collaboration associates with higher citation impact.

## Introduction

Research collaboration is believed to be beneficial for combining multiple skillsets and tackling applied problems with solutions that transcend disciplinary boundaries. Big science problems inherently require large numbers of researchers, and expensive equipment may require large consortiums to bid for its costs. Smaller groups of researchers may also benefit from sharing specialist tasks, skills or experience. Although collaboration might be advantageous in theory, there is mixed evidence about whether more collaborative scholars are more productive (using factional counting: Abramo, D'Angelo, & Murgia, 2017; Bidault & Hildebrand, 2014; Ductor, 2015; Lee & Bozeman, 2005). Collaboration is sometimes required by funding schemes and has support from numerous studies demonstrating that average citation impact tends to increase with the number of co-authors (see below). Solo research has not become extinct, however. Monographs are valued in the humanities and a degree of solo authorship seems to persist in all fields. The Science of Team Science (SciTS) field emerged in 2006 in response to the need to understand the complex factors that make collaborations successful (Hall, Vogel, Huang, Serrano, Rice, Tsakraklides, & Fiore, 2018), but the current paper focuses on international differences. These are important to understand so that collaboration-related science policies and strategies that work in one nation are not transferred to others where they fit less well.

There are many reasons why collaboration may vary internationally in its prevalence and impact. International differences in the importance of individualism in society

---

(Hofstede, 1980) may influence the extent to which researchers are inclined to collaborate. There are also country-specific funding regimes and research policies that may incentivise collaboration (Cao, Li, Li, & Liu, 2013; Tang, 2010), as well as differing national junior researcher mentoring strategies (e.g., Zhai, Su, & Ye, 2014). National challenges for scientific publication may also be helped by teamwork, such as language barriers (Brant & Rassouli, 2018; Duracinsky, Lalanne, Rous, Dara, Baudoin, Pellet, & Chassany, 2017; Tang, 2010). For example, a country in which English fluency is rare may regard literature reviewing and writing as a specialist task rather than a universal skill. Since geographical proximity (Kabo, Cotton-Nessler, Hwang, Levenstein, & Owen-Smith, 2014; Katz, 1994) and institutional support (e.g., Birnholtz, Guha, Yuan, Gay, & Heller, 2013) are important for collaborations, any national differences in the organisation or geography of universities may also influence collaboration strategies and effectiveness. Country size may also affect the likelihood of international collaborations (Ukrainski, Masso, & Kanep, 2014). Thus, the benefits of collaboration may vary by country and field.

Previous large-scale bibliometric research into collaboration has found that its prevalence has increased over time, that collaborative research attracts more citations than comparable solo research, and that team sizes are largest and increasing fastest in Science and Engineering, with Social Sciences second and Arts and Humanities last (Wuchty, Jones, & Uzzi, 2007). Collaboration 1900-2011 increased steadily across academia overall, split into two categories: Natural and Medical Sciences; and Social Sciences and Humanities (Larivière, Gingras, Sugimoto, & Tsou, 2015). In both areas, the field normalised citation impact of research 2005-2009 increased approximately logarithmically with the number of co-authors, partly due to additional self-citations for papers with more co-authors (Larivière, Gingras, Sugimoto, & Tsou, 2015). The citation advantage of extra co-authors has decreased over time and is greater when more countries are involved (Larivière, Gingras, Sugimoto, & Tsou, 2015). An investigation of Italian documents of multiple types in the Web of Science (WoS) core collection 2004-11 grouped them into 13 subject categories, with supplementary analyses of 217 WoS subject categories (some merged) (Abramo & D'Angelo, 2015). This study found some examples of fields where additional authors resulted in reduced field normalised citation impact, including Neurosciences (decreased impact from 2 to 3 authors) and Mathematics (decreased impact with more than 3 authors) categories, but this might be a statistical artefact given the relatively small sample sizes, the use of the arithmetic mean calculations and the large number of fields investigated (Abramo & D'Angelo, 2015). Similar results were found when journal impact was used instead of article citations.

The reason why collaborative papers tend to be more cited is not clear. In some fields, team authored work may not be higher quality (Bornmann, 2017). Whilst international collaboration is a strong indicator of higher citation impact (Didegah, & Thelwall, 2013; Van Raan, 1998), it does not tend to produce more novel research (Wagner, Whetsell, & Mukherjee, 2019), and in at least one field (biochemistry), international collaboration may not be beneficial unless the collaborator is from the USA (Sud & Thelwall, 2016). Thus, other factors are needed to explain the citation advantage of collaboration. As discussed above, self-citations are only a partial explanation. Larger and more diverse teams may attract more attention to the article through the authors' personal networks and specialisms (Larivière, Gingras, Sugimoto, & Tsou, 2015; Liao, 2010). People that collaborate tend to be more experienced (Van Rijnsoever & Hessels, 2011) and may therefore create more impactful research, although the outputs of older researchers may be less cited (Costas, Van Leeuwen, & Bordons, 2010). Another possibility is that collaborative research is more likely to be funded (Lee & Bozeman, 2005), and therefore tends to be better resourced. Funding does not seem to increase productivity (Defazio, Lockett, & Wright, 2009) but it seems to increase citation impact per publication (Jacob & Lefgren, 2011; Lewison, & Dawson, 1998; Thelwall, Kousha, Dinsmore, & Dolby, 2016). Despite these possibilities, it is not obvious whether collaboration should be less advantageous in any fields.

There are many different types of collaboration, at least as reflected by co-authorship. Each author typically contributes to a paper, with the first and last authors performing the most tasks (Larivière, Desrochers, Macaluso, Mongeon, Paul-Hus, & Sugimoto, 2016). A common type of collaboration is the PhD student/supervisor dyad, where the student performs most of the work and the supervisor directs or supports them. At the other extreme, huge long-term international projects may collaborate to address a shared complex challenge, such as for nuclear fusion experiments (Boisot, Nordberg, Yami, & Nicquevert, 2011). In between, there are collaborations of convenience, where two or more scholars combine to address a problem that they prefer to work on together, and collaborations of necessity, where multiple skills are necessary for a task. Moderate sized teams may also have members with "middle author" contributions that are not well defined (Mongeon, Smith, Joyal, & Larivière, 2017). A team may also form primarily to attract money, choosing partners based on funding council requirements. Finally, a report of US science and engineering performance showed that collaboration was more common for scientific publications with authors without academic affiliations (NSF, 2018), so collaborative research may be more industrially-focused.

Some science reports have examined collaboration from an international perspective. A UK-focused study of international collaboration confirmed that it associates with higher citation impact, also finding that a higher proportion of articles from the UK (51%) and France (51%) had an international collaborator than did articles from the other countries examined, including the USA, China and India. The UK had a higher proportion of solo articles (16% in 2010) than the other eleven countries examined, however, except for Russia (Figure 5.3 of Elsevier, 2017). A European-US comparison using Scopus data from 2011 found similar levels of solo authored articles (12% for the EU and 13% for the USA). It also found that EU-authored articles had greater increases in citation rates for collaboration of various types, such as inter-institutional and inter-regional co-authorship (Kamalski & Plume, 2013). The study most like the current paper investigated the field normalised citation impact of samples of articles from 2009 to 2015 (each year analysed separately) with 1-10 authors in selected Scopus categories for nine countries: Canada, China, France, Germany, Italy, Japan, Russian Federation, UK, USA (Thelwall & Sud, 2016). It found that citation impact tended to increase with the number of authors in all countries except the Russian Federation, where solo authored articles had more citation impact than small team collaborative articles. It did not examine the prevalence of collaboration and did not cover all of science, however. Other studies have also assessed the prevalence of international collaboration (Glänzel, 2001).

Despite extensive previous research into academic collaboration, there is a lack of science-wide complete comparisons of research collaboration within major research producing nations, with the partial exception of national prevalence reports and a previous sampling-based paper (Sud & Thelwall, 2016). This is an important omission because in the absence of this knowledge, it might be assumed that research collaboration tends to be internationally uniform, except for the Russian Federation exception previously identified.

This article therefore investigates international differences in the prevalence of collaboration and its citation impact. It also uses, for the second time, an impact comparison indicator designed to fairly represent the average citation impact of sets of journal articles, taking into account the skewed nature of citation data. The scope of this investigation is journal articles indexed by Scopus 2008-12 with the first author or all authors from any of the ten highest countries with the most publications in this period.
- RQ1: Are there international differences in the extent to which researchers collaboratively author journal articles? Does the answer change if only national collaborations are considered?
- RQ2: Are there any systematic country/field combinations for which collaboration does not associate with higher citation impact?

## Methods

The research design was to gather a large recent set of journal articles and to assess the field normalised citation impact advantage of international collaboration overall, when excluding international collaboration, and within broad fields.

Scopus was chosen for the data source since it has better coverage of non-English sources than the Web of Science (Mongeon & Paul-Hus, 2016) and at the time of writing seemed to have more reliable subject classifications than Microsoft Academic (Harzing & Alakangas, 2017) and Dimensions (Thelwall, 2018), with Google Scholar not offering subject-wide classifications (Martín-Martín, Orduna-Malea, Thelwall, & López-Cózar, 2018) and not usually permitting large scale downloading (Harzing, 1997). The data was collected in November 2018 as part of a generic dataset used for multiple studies. The years 2008-12 were chosen to give each article at least five years of citations, which should be adequate even in slow moving fields (Abramo, Cicero, & D'Angelo, 2011). Five publication years were used to increase the statistical power. A longer period would give more power but since collaboration has changed over time, lengthening the period also contaminates the data, and so five years is a compromise.

The ten countries with the most journal articles in Scopus 2008-12 were chosen fo for the maximum statistical power and relevance, whilst including differing research trajectories and geographic locations. Countries were attributed to articles through author affiliation data in Scopus. An article was classified as originating in a country if its first author's first affiliation was from that country. An article was classified as being authored solely from a country if at least one author had that country affiliation and no author had a different country affiliation. The focus of the analysis is on national collaboration so that patterns are less affected by international considerations. Nevertheless, national collaborations can have hidden international dimensions if one of the participants has worked abroad (Jonkers & Tijssen, 2008), and international collaboration can be with diaspora scientists associated with the collaborating country (Wang, Xu, Wang, Peng, & Wang, 2013). National collaboration in large countries, such as China, India, and the USA, can also have characteristics of international collaboration if it is between different regions (Sun & Cao, 2015).

Articles were grouped using the Scopus broad subjects (www.elsevier.com/solutions/scopus/how-scopus-works/content). They were categorised for collaboration using the Scopus author IDs to count the number of distinct authors. These IDs are algorithmically derived and are likely to be less accurate for countries like China with very common names (in the Latin alphabet), but they should be reasonably accurate within articles since by-lines normally indicate multiple affiliations for single authors.

The average citation impact of a group of articles was calculated with the Mean Normalised Log-transformed Citation Score (MNLCS) (Thelwall, 2017). Field normalised indicators (Waltman, van Eck, van Leeuwen, Visser, & van Raan, 2011) allow citation impact to be compared between articles from different fields and years, despite both affecting average citation counts, by dividing each citation count by the average for the field(s) and year in which it was published. Irrespective of field and year, a score above 1 always equates to research with more citations that the world average for the field and year of publication. The MNLCS log transforms citations with $ln(1 + c)$ before performing the calculations because sets of citation counts are highly skewed and the arithmetic mean of a skewed set of numbers can be dominated by a few highly cited articles.

The results are reported separately for team sizes 1 to 25 and the remainder are combined into a 25+ authors set. Since there are relatively few team sizes larger than 25 (and much smaller team sizes for some countries and fields), the citation results are not reliable for larger teams. The cut-off 25 was chosen heuristically.

It is impossible to give a statistically significant answer to the field comparison aspect of RQ2 because of the many possible differences to compare (e.g., 10 countries x 27 fields x 26 author number groupings) and so differences will be identified and presented for qualitative interpretation instead.

## Results

The results are summarised as graphs, with key points and differences discussed. All figures are available in a spreadsheet in the online supplement (10.6084/m9.figshare.9038963), together with 95% confidence intervals, so that individual countries and disciplines can be examined if they are not identifiable in the graphs, or if they cover fields that are not included below. Australia is also covered by the online data.

### *Research team size*

There are substantial differences in the extent to which the ten selected countries collaborated nationally (Figure 1) and overall (Figure 2). The clearest contrast in the graph is perhaps between the UK/USA and China/Japan for national collaboration because the modal number of UK authors is 1, whereas it is 4 for China and Japan. Few (4%) Chinese national papers have only one author. Overall, however, Japan has a slightly larger average team size than China (Figure 3) because it has more large teams.

It might not be a coincidence that the three countries that collaborate least overall (Figure 3) are majority native English speaking, although there is no causal evidence for language being a factor in collaboration.

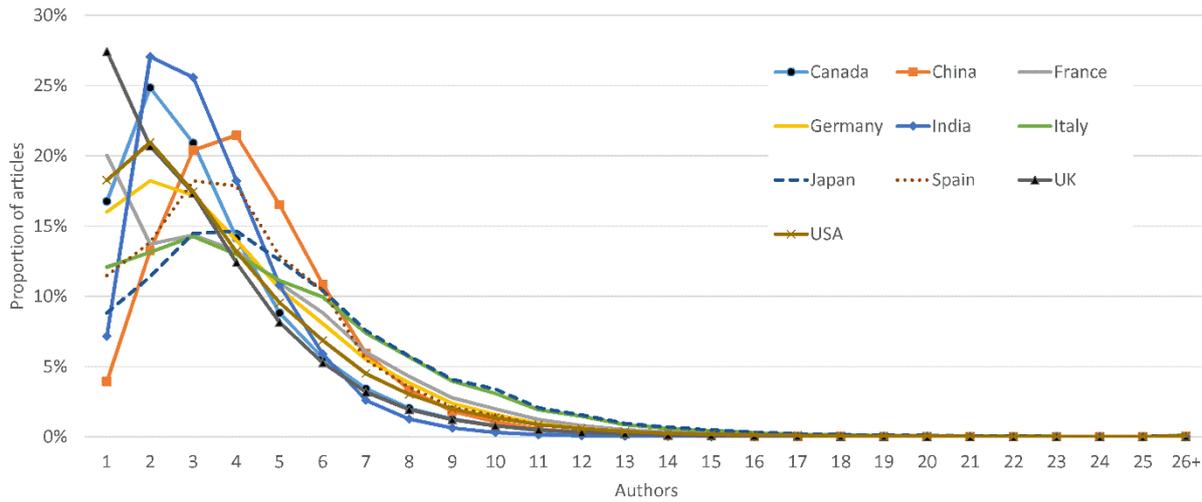

Figure 1. The proportion of national research articles with a given number of authors, by country. Data: journal articles in Scopus 2008-12. Confidence intervals are available in the online materials.

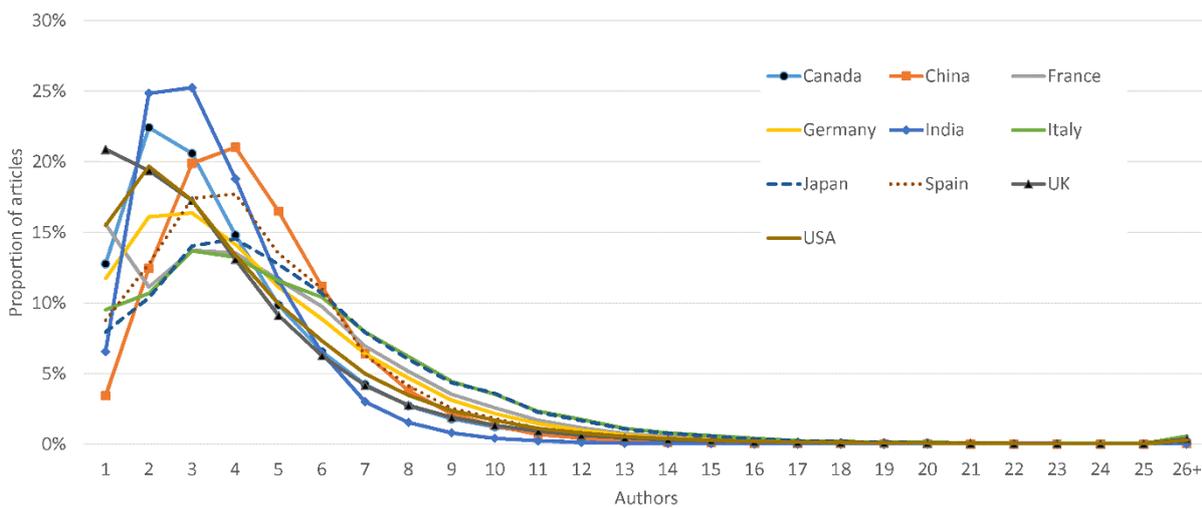

Figure 2. The proportion of all (national and international) research articles with a given number of authors, by country. Data: journal articles in Scopus 2008-12.

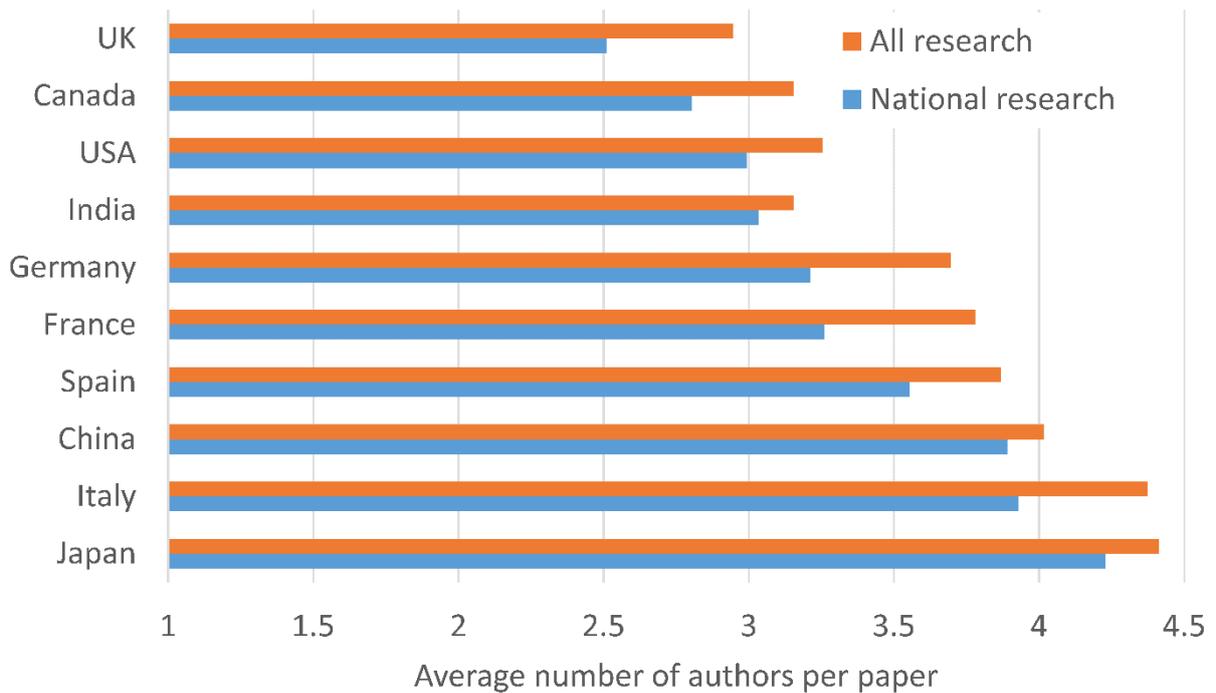

Figure 3. The average (geometric mean) number of authors per article, by country. Data: journal articles in Scopus 2008-12.

There are substantial differences between fields in the extent of collaboration 2008-12. This pattern persists for national research from every country examined. For example, it applies to the UK (Figure 4), which has the least collaboration overall, and to Japan (Figure 5), which has the most. Thus, the difference between the UK and Japan overall for collaboration is not that Japan researches more in collaborative fields but that researchers with Japanese affiliations are more likely to collaborate, whichever field they work in.

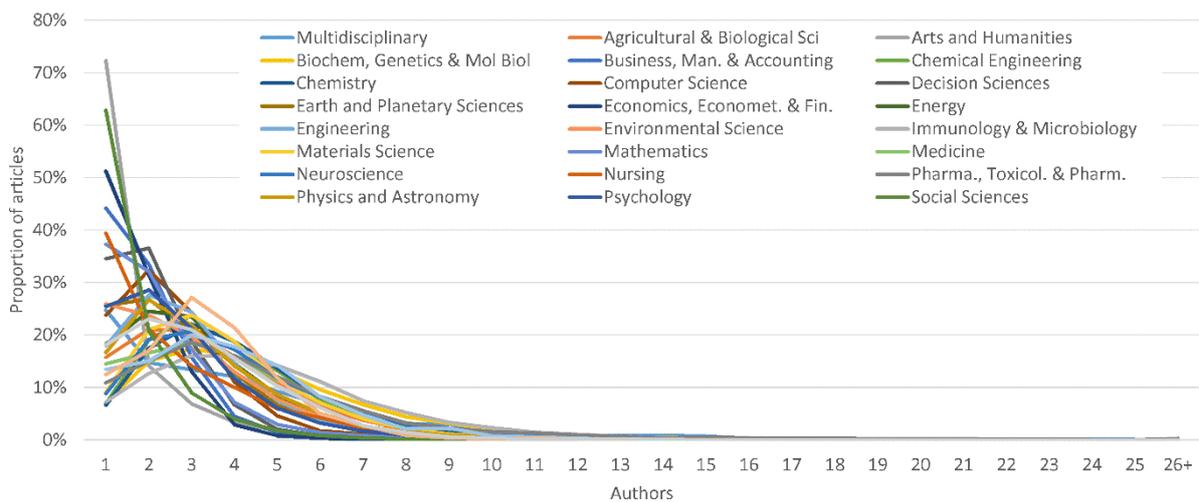

Figure 4. The proportion of national UK research articles by number of authors, for each Scopus broad research field. Data: journal articles in Scopus 2008-12.

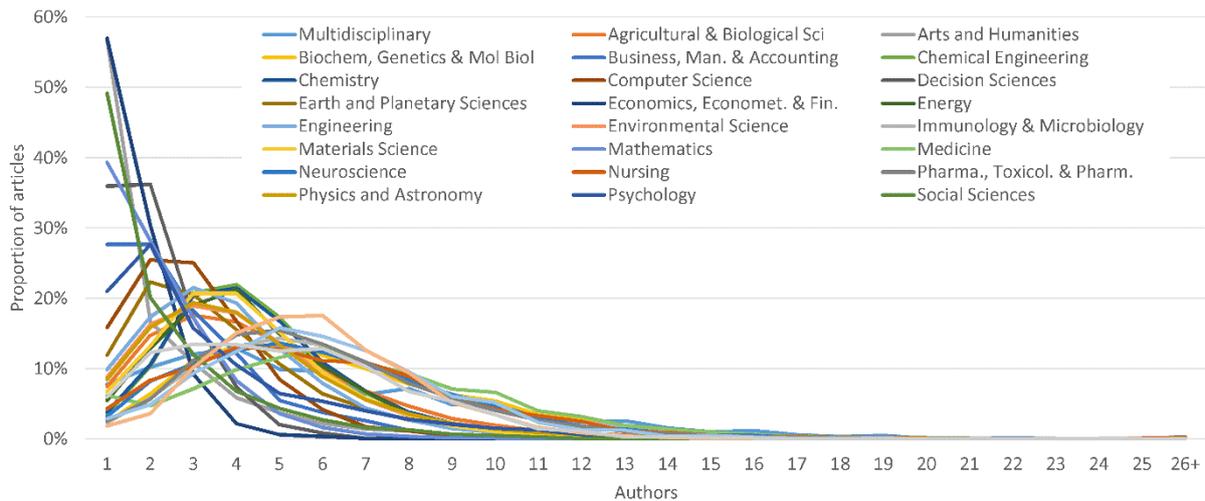

Figure 5. The proportion of national Japanese research articles by number of authors, for each Scopus broad research field. Data: journal articles in Scopus 2008-12.

The important and relatively collaborative field of Medicine (Figure 6) illustrates substantial international variations in the extent of collaboration within a field. The modal number of authors is 3 for the USA, UK and Canada but 6 for Italy, Japan, and Spain.

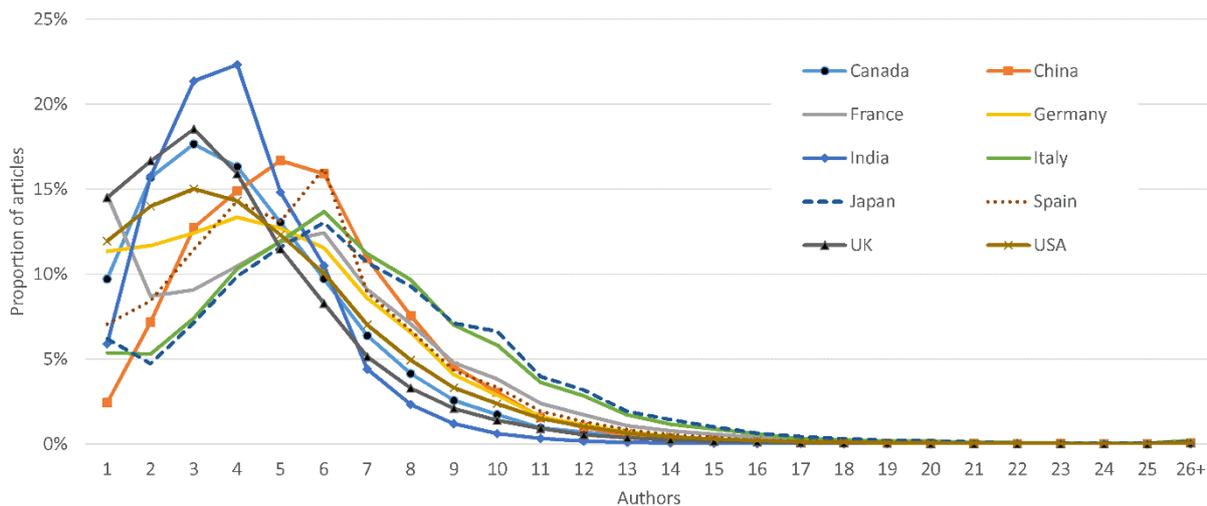

Figure 6. The proportion of national research articles from Medicine by number of authors, for each country. Data: journal articles in Scopus 2008-12.

Solo authorship is modal for all ten countries in the Social Sciences (Figure 7) but with greatly varying overall proportions. China is an outlier for the far greater prevalence of collaboration.

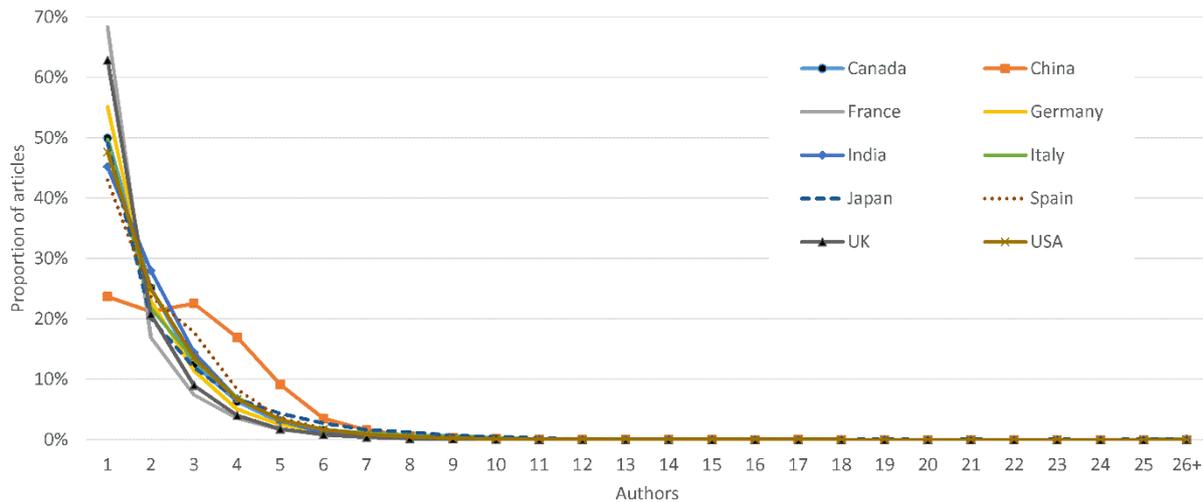

Figure 7. The proportion of national research articles from Social Science by number of authors, for each country. Data: journal articles in Scopus 2008-12.

Biochemistry and Molecular Biology (Figure 8) and Engineering (Figure 9, see also Figure 6) illustrate the similarity of country shape differences between fields. The latter graph is close to a squashed version of the former, although there are some other differences. For example, Canada and India largely overlap in Figure 9 but India is higher for 2-4 authors in Figure 8 and more different in Figure 6. Figures for other fields are available online.

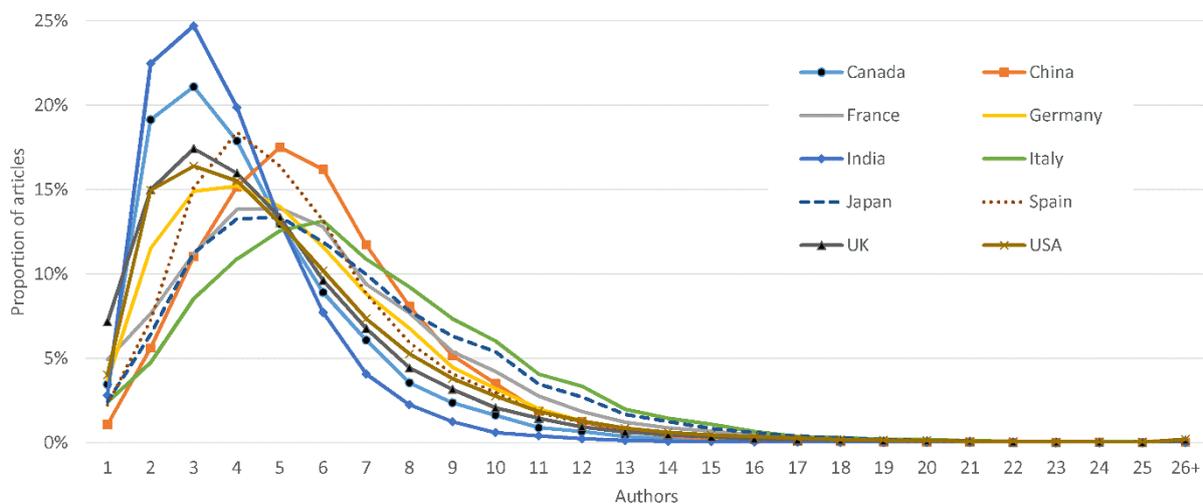

Figure 8. The proportion of national research articles from Biochemistry and Molecular Biology by number of authors, for each country. Data: journal articles in Scopus 2008-12.

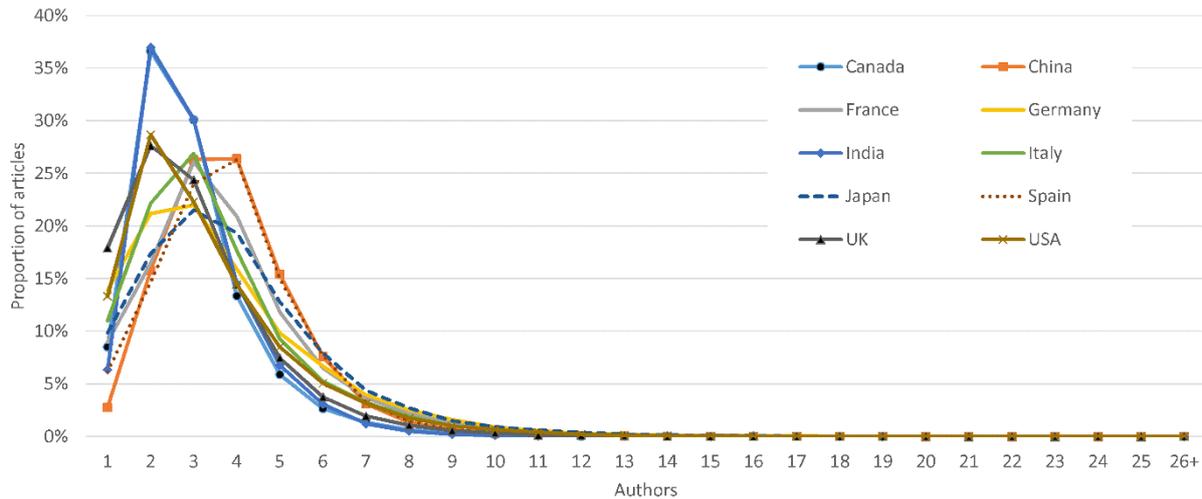

Figure 9. The proportion of national research articles from Engineering by number of authors, for each country. Data: journal articles in Scopus 2008-12.

## Research team size and citation impact

For all countries except China, there is a sharp jump in average citation impact from solo authored articles to articles with 2 authors, followed by an approximately linear increase in citation impact with extra authors, but with the magnitude of the increase probably declining (Figure 10). Low numbers make the pattern unreliable after about 11 authors. China has approximately linearly increasing citation impact with authors up to about 12, then with additional authors seeming to generate less increase in citation impact.

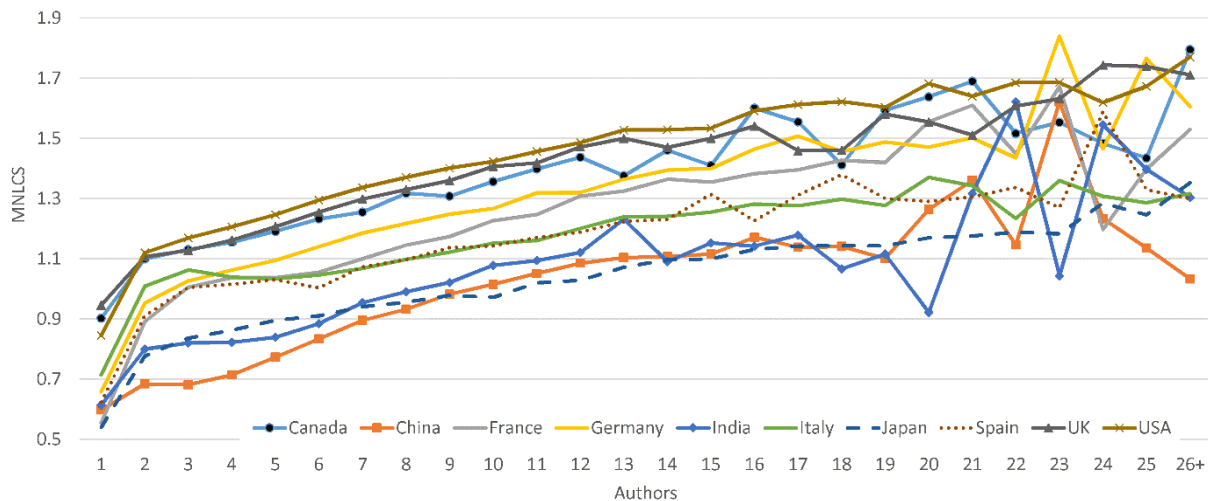

Figure 10. The average field normalised citation impact (MNLCS) of national research articles by number of authors, for each country. Data: journal articles in Scopus 2008-12.

Citation impact graphs for individual fields have much less precise lines due to substantially less underlying data. Although not statistically significant, both Computer Science for China (Figure 11) and Business, Management & Accounting for China and Japan (Figure 12), illustrate that collaboration may not always strongly associate with higher impact. For Germany, the lower citation impact in Business, Management & Accounting for articles with 4 or 5 authors than for articles with 3 authors (Figure 12) is statistically significant (non-overlapping confidence intervals, as shown in the supplementary material) from the perspective of a single test, but some statistically significant results are to be expected when large numbers of comparisons are made (Perneger, 1998), so it is not safe to conclude that the difference is statistically significant.

To check for statistical significance for the key case, the same test was repeated for Germany only for *Business, Management & Accounting* in two previous non-overlapping five-year periods, 2003-07 and for 1998-2002. For *Business, Management & Accounting* 2003-07 in Germany, the citation impact of 5 authors was statistically significantly lower than that of 3 authors, agreeing with the 2008-12 data, but the citation impact of 4 authors was higher than that of 3 authors, disagreeing with the 2008-12 data. For *Business, Management & Accounting* 1998-2002 in Germany, the citation impact of 5 and 4 authors were not statistically significantly lower than that of 3 authors (one was higher, one was lower), again partially disagreeing with the 2008-12 data. Although the value of business-related collaboration in Germany may have changed over time, these extra tests tend to confirm that it is not safe to conclude that individual country/discipline combinations have anomalous relationships between collaboration and impact based on individual statistical tests. The anomalous average impact differences between team sizes are not large enough to be statistically significant after a familywise error rate correction for multiple tests (e.g., Bonferroni).

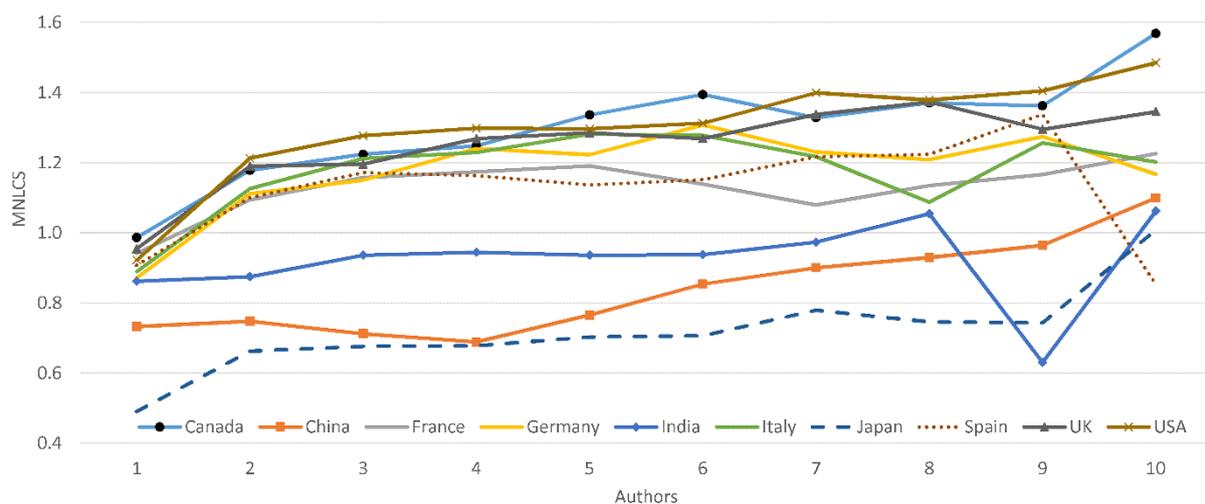

Figure 11. The average field normalised citation impact (MNLCS) of national research articles in *Computer Science* by number of authors, for each country. Data: journal articles in Scopus 2008-12.

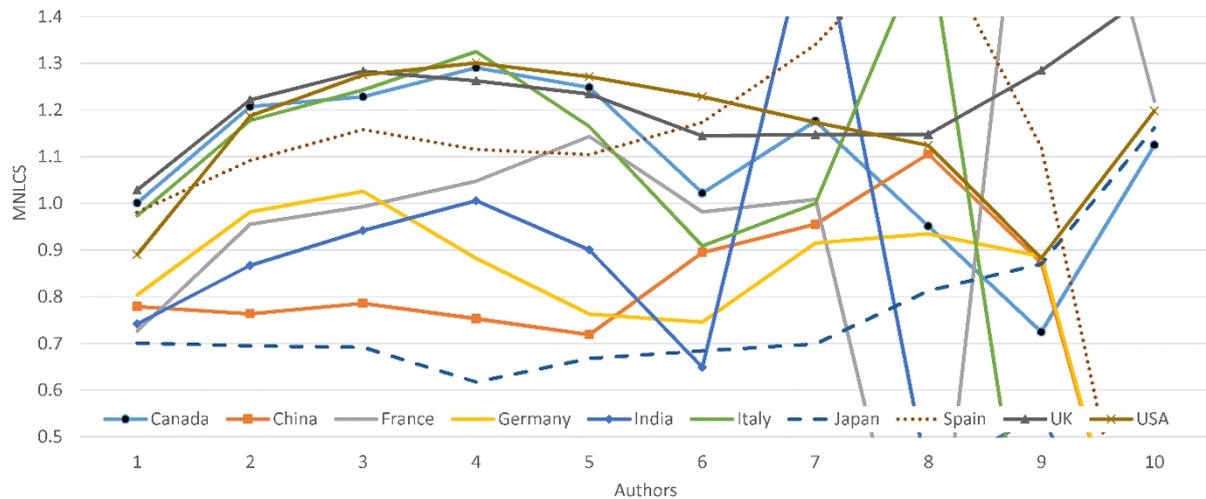

Figure 12. The average field normalised citation impact (MNLCS) of national research articles in *Business, Management & Accounting* by number of authors, for each country. Data: journal articles in Scopus 2008-12.

## Discussion

This study has several limitations. There may be important differences in collaboration patterns within regions in large countries (Sun & Cao, 2015; Tang & Shapira, 2010) that will not be reflected. The use of Scopus is also a limitation because of its dominance by English-language sources. It is possible, for example, that Chinese solo-authored publications would be more likely to be published in Chinese-language journals and less likely to be indexed in Scopus as a result.

The results show substantial international variations in the extent of co-authorship in research publishing, whether including or excluding international teams. These differences occur within broad fields and so cannot be accounted for by international variations in research specialisms. Whilst the UK is the least likely of the ten countries studied to collaborate, Japan, Italy and China have the largest average team sizes. The results have a partial match with the Hofstede (1980, 2011) cultural dimension of Individualism. People in countries with a high degree of individualism are less likely to think in collective terms and value communal goals. Although the least individualist cultures collaborate most (China, and, to a lesser extent, Japan and Spain), Italy is an important exception for its high degree of individualism and extensive collaboration, and Japan is only moderately individualist. The case of Italy might be related to the Italian research assessment exercise (Franceschini & Maisano, 2017), where the importance of citations might lead academics to collaborate more. Overall, however, the lack of a close alignment with the Hofstede Individualism dimension suggests that research-specific factors are more important than general societal cultures. Thus, the international differences in the extent of collaboration found here lack a simple explanation.

The association found between collaboration and higher average citation impact confirms many prior studies (Larivière, Gingras, Sugimoto, & Tsou, 2015; Thelwall & Sud, 2016). The results extend prior knowledge by finding that this occurs when an appropriate averaging measure is used science-wide, so that the field normalised citation impact scores cannot be dominated by a minority of individual highly cited articles. They also extend it by reporting the effect in terms of the number of authors rather than the type of collaboration (e.g., national or international).

The citation results also confirm a previous comparison of the USA and Europe (Kamalski & Plume, 2013) and a comparison of nine countries (eight of which are in the current paper) (Thelwall & Sud, 2016) that the benefits of collaboration vary internationally rather than being constant. The results extend the previous study by finding differences for two extra countries (India, Spain) and with more robust evidence from a science-wide rather than sampled data set. The results also extend the previous findings with the discovery that in China the citation impact benefit of a second author is minor. The current study did not find a citation disadvantage for collaboration comparable to that previously found for the Russian Federation (Thelwall & Sud, 2016), however.

The results support a previous finding that collaboration does not confer a citation advantage in all country/discipline combinations (Italy: Abramo, & D'Angelo, 2015; multiple large countries: Thelwall & Sud, 2016) with additional examples with larger data sets. Nevertheless, this remains a tentative finding given the likelihood of natural statistical variations in citation impact due to the relatively low sample sizes involved and the numerous country/field/author count combinations that could be compared.

The differing international rates of co-authorship do not necessarily directly reflect differing rates of research collaboration. It is likely that there are international and field differences in the extent to which contributions are acknowledged by co-authorship, but it seems unlikely that such differences would be large enough to account for the results found here.

## Conclusions

There are substantial international variations in the extent to which countries co-author Scopus-indexed articles. These need further exploration to discover the underlying reasons for the differences. Similarly, there are substantial international variations in the extent to which collaboration associates with higher citation impact and again this does not have a clear explanation.

If the larger average team size in some countries is due to a need for English language skills as an additional requirement for international publishing, then this would explain the larger team sizes. If, on the other hand, co-authorships are given for lesser contributions in some countries then this would have implications for research ethics (COPE, 2014). It would also have implications for research evaluation for individual academics because some academics would have longer publication lists due to more lenient authorship practices.

Given the international differences in collaboration found, research funders and managers should be careful to avoid assuming that collaboration strategies that are effective in one nation would easily transfer to another. Similarly, scientists collaborating with researchers from another country should be prepared to accommodate differing team sizes in their projects. Moreover, given that collaboration does not always associate with higher research impact, national research managers and funders should be sensitive to the possibility that it should not be promoted in some fields.